\documentclass[aps,prd,11pt,notitlepage,nofootinbib,superscriptaddress,showkeys,showpacs]{revtex4-1}
\usepackage{amsmath,amssymb,amsthm,latexsym}
\usepackage{stmaryrd,wasysym,upgreek,mathrsfs,dsfont}
\usepackage[english]{babel}
\usepackage{graphicx,color}
\usepackage{xspace}
\usepackage{graphicx}
\usepackage{pifont,dsfont}
\usepackage{marvosym}

\newcommand{\be}{\begin{equation}}
\newcommand{\ee}{\end{equation}}
\newcommand{\bqa}{\begin{eqnarray}}
\newcommand{\eqa}{\end{eqnarray}}
\newcommand{\bea}{\begin{eqnarray}}
\newcommand{\eea}{\end{eqnarray}}

\newcommand{\cF}{{\cal F}}

\newcommand{\cG}{{\cal G}}
\newcommand{\cB}{{\cal B}}

\DeclareMathOperator{\Tr}{Tr}

\begin{document}

\title{\Large \bf The Double Scaling Limit in Arbitrary Dimensions: A Toy Model}

\author{{\bf Razvan Gurau}}\email{rgurau@perimeterinstitute.ca}
\affiliation{Perimeter Institute for Theoretical Physics, 31 Caroline St. N, ON N2L 2Y5, Waterloo, Canada}

\begin{abstract}
Colored tensor models generalize matrix models in arbitrary dimensions yielding a statistical theory of random 
higher dimensional topological spaces. They admit a $1/N$ expansion dominated by graphs of spherical topology. 
The simplest tensor model one can consider 
maps onto a rectangular matrix model with skewed scalings. We analyze this simplest toy model and show 
that it exhibits a family of multi critical points and a novel double scaling limit. We show in $D=3$ dimensions 
that only graphs representing spheres contribute in the double scaling limit, and argue that similar results hold
for any dimension. 
\end{abstract}

\medskip


\maketitle

\section{Introduction}

Random matrix models  \cite{Di Francesco:1993nw} provide a statistical theory of random  
discretized Riemann surfaces. The amplitudes of the ribbon Feynman graphs of their perturbative expansion 
support a $1/N$ expansion 
\cite{'tHooft:1973jz, Brezin:1977sv} (where $N$ is the size of the matrices) indexed by the genus of the surfaces.
In the large $N$ limit the planar graphs corresponding to surfaces of spherical topology dominate \cite{Kazakov:1985ds,mm}. 
Statistical models of fluctuating geometry can be thought as providing either a regularization 
or a fundamental description of quantum gravity \cite{ambjorn-book}. The large-$N$ limit of matrix models 
yields in two dimensions an analytic description of dynamical triangulations 
\cite{david-revueDT,ambjorn-houches94},  whose link to non-critical 
string theory in the continuum limit is well-understood \cite{Di Francesco:1993nw}. 
Higher-dimensional models of dynamical triangulations have not been equally successful in providing a sensible continuum limit for quantum 
gravity \cite{Loll:1998aj,Ambjorn:1995dj}, although a non-local modification, which goes under the name of causal dynamical triangulations 
\cite{Ambjorn:2010ce} has produced substantial evidence for the emergence of an extended geometry at large scale 
\cite{Ambjorn:2005qt,Ambjorn:2011ph,Ambjorn:2011cg}.

The family of planar graphs \cite{Brezin:1977sv,Kazakov:1985ds,mm} dominating the large $N$ limit of matrix models 
is summable with finite radius of convergence. When the coupling constant approaches a critical value $g_c$, 
the free energy is dominated by graphs with a large number of vertices and exhibits a critical behavior. It is in this regime
that the system reaches its continuum limit and the critical exponents can be evaluated.

More general matrix models exhibit a complex critical behavior, specific choices of the potential leading to 
multi critical points \cite{Kazakovmulticrit}. They correspond to $(m+1,m)$ conformal matter coupled
to Liouville gravity  \cite{Kazakovmulticrit}, and map on conformal field theories on fixed geometries via the KPZ correspondence 
\cite{Knizhnik:1988ak, david2, DK,Dup}. The contributions to the free energy (that is the partition function for 
connected surfaces) at fixed genus exhibit a 
critical behavior when the coupling $g$ goes to the same fixed $g_c$.
The corresponding critical exponents are such that the double scaling limit $N\to \infty$, $g\to g_c$, 
with $ N(g-g_c)^{1 + \frac{1}{2m} } =\kappa^{-1} $ fixed combines all genera \cite{double,double1,double2} and leads to a 
well defined continuum theory with fixed Newton's constant \cite{Di Francesco:1993nw}.

Random matrices generalize in higher dimensions 
to random tensors \cite{ambj3dqg,sasa1,mmgravity}, whose perturbative expansion performs a sum over random higher dimensional geometries.  
Although some power counting estimates have been obtained \cite{pcont1,pcont2,pcont3,pcont4,pcont5,pcont6,pcont7},
and the symmetries of tensor models could be analyzed \cite{Sasa2,Sasa3,Sasa4}
the key to analytical rather than numerical results, namely the $1/N$ expansion, was missing until recently.
That situation has changed with the discovery of such a $1/N$ expansion \cite{Gur3,GurRiv,Gur4} for {\it colored} 
\cite{color,PolyColor,lost,orie} random tensors. The amplitude of their graphs supports a $1/N$ expansion indexed by
the {\it degree}, a positive integer, which plays in higher dimensions the role the genus played in two dimensions. 
The leading order graphs, baptized melonic \cite{Bonzom:2011zz}, triangulate the $D$-dimensional sphere and  
form a summable series. When the coupling constant approaches its critical value, the free energy exhibits a critical 
behavior and,  like in matrix models, the colored tensor models reach their continuum limit
dominated by triangulations with an infinite number of simplices.  The entropy exponent of the melonic series, analogous to the string 
susceptibility $\gamma_{\rm string} = -1/2$ of the 1-matrix model for the pure gravity universal class, is 
$\gamma_{\rm melons} = 1/2$, \cite{Bonzom:2011zz}.
This discovery led to the possibility of new analytical  investigations of dynamical triangulations models and their continuum limit 
in $D\geq 3$ dimensions \cite{Bonzom:2011zz,Gurau:2011tj}.
Colored random tensors are a promising tractable discretization of quantum gravity in three and more dimensions and the subject is 
developing fast \cite{sym4,class1,Baratin:2011tg,Oriti:2010hg,Geloun:2011cy}. 
The understanding of the leading (melonic) order of colored tensor models allows the study of the coupling of statistical systems 
to random geometries in arbitrary
dimension. Thus one can prove that, unlike in two dimensions \cite{Ising2}, the Ising model on a random lattice in higher dimensions 
does not exhibit a phase transition \cite{IsingD},
while the dually weighted models (introduced in \cite{dual2} in two dimensions) counting dynamical triangulations with a non trivial measure 
factor, do \cite{EDT}. A detailed introduction and review of colored tensor models is \cite{coloredreview}. 

The Schwinger Dyson equations (SDE) of matrix models translate into constrains
(satisfied by the partition function) which obey the Virasoro algebra \cite{Ambjorn:1990ji,Fukuma:1990jw,Makeenko:1991ry} . 
A similar line of inquiry can be pursued in higher dimensions by integrating all colors but one in a colored tensor model
and attributing an independent coupling constant to each effective vertex. The algebra of constrains is, unsurprisingly, much
more involved, but still manageable at leading order \cite{Gurau:2011tj}. 
The detailed study of this algebra will lead to a precise understanding of the various critical behaviors and 
a classification of the continuum limits of colored tensor models. 

In its full generality this classification is for now out of reach and one must must content with the study of some simplified
toy models where a more detailed analysis can be pursued. This is done in the present paper by restricting 
to a sub sector of the tensor models consisting in a matrix model with skewed scalings. A tensor $T_{n^1\dots n^D}$ 
with $D$ indices can be seen as a $N \times N^{D-1}$ matrix $T_{n^1 \vec n}$, with $\vec n = n^2\dots n^D$. 
General tensor interactions correspond to arbitrary contractions of indices, hence they do not respect this
splitting. In the sequel we restrict to the subclass on interactions which do, that is we only take into account
interactions of the form $\Tr [( T T^{\dagger} )^p]$. In order to obtain a sensible theory one needs to adapt
appropriately the scalings of the terms with the large parameter $N$. This toy model is natural once one 
takes into account the full SDE's of colored tensor model: it consist in restricting to a model whose leading 
order SDE's close a Virasoro algebra. The matrix model with skewed scalings can also be interpreted as a 
model of random surfaces with patches. 
 
Using techniques consecrated for matrix models, but taking into account the effect of the unusual 
scaling we present in this paper the full solution of the toy model via the all orders SDE's.
We generalize for all $D$ the multi critical points of matrix models \cite{Kazakovmulticrit}, and
find that their susceptibility exponents are $1-\frac{1}{m}$,
identical with those of multi critical polymers \cite{Ambjorn:1990wp}.
We the particularize for convenience to $D=3$ and show that the model admits a double scaling limit
$N\to \infty$ $g\to g_c$ with $N(g-g_c)^{1+\frac{1}{m}} = \kappa^{-1}$ fixed.
We then go on to show that the double scaling limit we uncover is {\it not} a summation over topologies: indeed 
{\it only planar graphs} with trivial topology contribute to the double scaling limit. Viewed as graphs of a 
colored tensor model in $D=3$, these graphs always represent spheres. We argue that similar results hold in all dimensions.

This paper is organized as follows. In section \ref{sec:model} we present the colored tensor model and its relation to the 
toy model of a matrix with skewed scalings. In section \ref{sec:mmodel} we analyze the toy model and
its multi critical points. In section \ref{sec:SDE} we deduce the all orders SDEs
and present their iterative solution. In section \ref{sec:multicrit} we present the double scaling limit
of our toy model in $D=3$, and in section \ref{sec:conc} we discuss the implications of these results
for general tensor models.

\section{From Colored Tensor Models to a Matrix Model With Skewed Scalings} \label{sec:model}

Our starting point is the independent identically distributed colored tensor model with one coupling. 
We denote $\vec n_i$, for $i=0,\dotsc,D$, the $D$-tuple of integers
 $\vec n_i = (n_{ii-1},\dotsc, n_{i0},\; n_{iD}, \dotsc, n_{ii+1}) $, with
$n_{ik}=1,\dotsc, N$. This $N$ is the size of the tensors and the large $N$ limit defined in 
\cite{Gur3,GurRiv,Gur4} represents the limit of infinite size tensors. 
We set $n_{ij} = n_{ji}$. Let $\bar \psi^i_{\vec n_i},\; \psi^i_{\vec n_i}$, with $i=0,\dotsc, D$, be $D+1$ couples of complex
conjugated tensors with $D$ indices. The independent identically distributed (i.i.d.) colored tensor
model in dimension $D$ \cite{coloredreview} is defined by the partition function
\begin{gather}
\nonumber e^{ - N^D F_N(\lambda,\bar\lambda)} = Z_N(\lambda, \bar{\lambda}) = \int \, d\bar \psi \, d \psi
\ e^{- S (\psi,\bar\psi)} \; , \\
S (\psi,\bar\psi) = \sum_{i=0}^{D} \sum_{ n} \bar \psi^i_{\vec n_i} \psi^i_{\vec n_i}  +
\frac{\lambda}{ N^{D(D-1)/4} } \sum_{ n} \prod_{i=0}^D \psi^i_{ \vec n_i } +
\frac{\bar \lambda}{ N^{D(D-1)/4} } \sum_{ n}
\prod_{i=0}^D \bar \psi^i_{ \vec n_i } \; . \label{eq:iid}
\end{gather}
$\sum_{ n}$ denotes the sum over all indices $ n_{ij}$ from $1$ to $N$.
The tensor indices $n_{ij}$ need not be simple integers (they can for instance index the
Fourier modes of an arbitrary compact Lie group, or even of a finite group of large 
order \cite{bahr-dittrich-ryan:baby}). 
The partition function of equation \eqref{eq:iid} is evaluated by colored Feynman graphs
\cite{color,lost,PolyColor}. The tensors have {\it no} symmetry properties under permutations 
of their indices (i.e. all $\psi^i_{\vec n_i}, \bar \psi^i_{\vec n_i}$ are independent).
The colors $i$ of the fields $\psi^i, \bar{\psi}^i$ induce important
restrictions on the combinatorics of the graphs. They have two types of vertices,
say one of positive (involving $\psi$) and one of negative (involving $\bar \psi$).
The lines always join a $\psi^i$ to a $\bar{\psi}^i$ and possess a color index, $i$.
Any Feynman graph $\cG$ of this model is an orientable simplicial pseudo manifold \cite{lost,orie} and
the colored tensor models provide a statistical theory of random triangulations in 
$D$ dimensions, generalizing random matrix models.

\begin{figure}[htb]
\begin{center}
 \includegraphics[width=3cm]{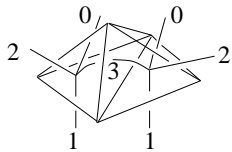}  
\caption{The dual complex in $D=3$.}
\label{fig:complex}
\end{center}
\end{figure}

One can picture the topological space associated to a graph in a rather simple manner.
The vertices of the graph correspond to the $D$ simplices of the simplicial complex. The halflines of a vertex represent 
the $D-1$ simplices bounding a $D$ simplex and have a color. Any lower dimensional subsimplex is colored by the colors of 
the $D-1$ simplices sharing it. 
In figure \ref{fig:complex} we sketched the dual complex in $D=3$ dimensions. The vertices are
dual to tetrahedra. A triangle (say $3$) is dual to a line (of color $3$) and separates two tetrahedra.
An edge (say common to the triangles $2$ and $3$) is dual to a face (and indexed by two colors $2$
and $3$). A vertex (say common to the triangles $0$, $2$ and $3$) is indexed by the three colors
$0$, $2$ and $3$.

The {\bf $n$-bubbles} of the graph  are the maximally connected subgraphs made of lines with $n$ fixed colors.
They are associated to the $D-n$ simplices of the pseudo manifold.
For instance, the $D$-bubbles are the maximally connected subgraphs containing
all but one of the colors. They are associated to the $0$ simplices (vertices) of the pseudo-manifold.

The tensor indices $n_{jk}$ are preserved along the faces of the graph (which are readily identified as 
bi colored connected subgraphs). 
The amplitude of a graph with $2p$ 
vertices and $\cF$ faces is \cite{Gur4}
\bea\label{eq:ampli}
A(\cG)= (\lambda\bar\lambda)^p N^{-p \frac{D(D-1)}{2}+ \cF } \; .
\eea

We generalize the colored tensor model with one coupling to a model 
with an infinity of couplings. First we integrate all colors save one, and second we ``free'' the couplings of the operators 
in the effective action for the last color. When integrating all colors save one the partition function becomes
\bea
&& Z = \int d\psi^0 d\bar \psi^0 \; e^{-S^0(\psi^0, \bar \psi^0)} \crcr
&& S^0(\psi^0, \bar \psi^0) =  \sum \bar \psi^0_{ \vec n_0 } \psi^0_{\vec n_0 }
 + \sum_{ \cB^{\widehat{0}} } \frac{(\lambda\bar\lambda)^{p} }{ \text{Sym}(\cB^{\widehat{0}}) } 
\; \Tr_{\cB^{\widehat{0}}} [\bar\psi^0, \psi^0 ]
\;  N^{-\frac{D(D-1)}{2}p + \cF_{\cB^{\widehat{0}} } }
\eea
where the sum over $\cB^{\widehat{0}}$ runs over all connected vacuum graphs with colors $1,\dots D$ 
(i.e. over all the possible $D$-bubbles with colors $1,\dots D$) and $p$ vertices.
The operators $\text{Tr}_{\cB^{\widehat{0}} } [\bar\psi^0, \psi^0 ] $ in the effective action for the last color
are tensor network operators. Every vertex of $\cB^{\widehat{0}} $ is decorated by a tensor
$\psi^0_{\vec n_0}$ or $\bar \psi^0_{ \vec {\bar n}_0 }$, and the tensor indices are contracted 
as dictated by the graph $\cB^{\widehat{0}} $. 
We denote $v$, $\bar v$ the positive (resp. negative) vertices of $\cB^{\widehat{0}} $, 
and $l^i_{v\bar v}$ the lines (of color $i$)
connecting the positive vertex $v$ with the negative vertex $\bar v$.
The operators write
\bea\label{eq:tensnet}
 \text{Tr}_{\cB^{\widehat{0}}} [\bar\psi^0, \psi^0 ] = \sum_{n}
 \Bigl( \prod_{v,\bar v \in \cB^{\widehat{0}}} \bar \psi^0_{\vec {\bar  n}^{\bar v}_0 } \psi^0 _{\vec n^v_0 } \Bigr)
 \Bigl( \prod_{i=0}^{D-1} \prod_{ l^i_{v\bar v} \in \cB^{\widehat{ 0 }}  } 
\delta_{n^{v }_{0i} {\bar n}^{\bar v }_{0i} } \Bigr) \; ,
\eea
where all indices $n$ are summed. Note that, as all vertices in the bubble belong to an unique line of a given color, all
the indices of the tensors are paired. The scaling with $N$ of an operator can be evaluated 
and the effective action for the last color writes (dropping the index $0$)
\bea
 S^D ( \psi , \bar \psi) =  \sum \bar \psi_{ \vec  n } \; \psi_{  \vec n }
 + N^{D-1} \sum_{ \cB }  \frac{ (\lambda \bar\lambda)^{ p  } }{ \text{Sym}( \cB  )} 
\; N^{-(D-1)p -\frac{2}{(D-2)!}\omega(\cB) } \;
\text{Tr}_{ \cB  }  [ \bar\psi  , \psi ] \; ,
\eea
with $\omega(\cB)$ a non negative integer \cite{Gurau:2011tj,coloredreview}.
Attributing to each operator its coupling constant and rescaling the field to $T = \psi N^{-\frac{D-1}{2}}$, we
obtain the partition function of colored tensor model with generic potential 
\bea\label{eq:genmod}
&& Z =e^{-N^D F(t_{\cB } ) }= \int d\bar T dT \; e^{-N^{D-1} S(\bar T, T)} \; ,\crcr
&& S (\bar T, T ) = \sum \bar T_{\vec n} T_{\vec n} +   \sum_{ \cB  }  t_{ \cB  } \quad  
N^{ -\frac{2}{(D-2)!}\omega(\cB) } \; \text{Tr}_{ \cB } [ \bar T ,  T ] \; .
\eea

Although in the end we deal with an unique tensor $T$, the colors are crucial for the
definition of the tensor network operators in the effective action. The initial vertex of the tensor model
described a $D$ simplex. The tensor network operators describe (colored) polytopes in $D$ dimensions obtained by 
gluing simplices along all save one of their boundary $D-1$ simplices around a point (dual to the bubble $\cB $).
This is in strict parallel with matrix models, where higher degree interactions represent polygons obtained by
gluing triangles around a vertex. Each index of a tensor $T$ inherits a color. 

\begin{figure}[htb]
 \includegraphics[width=6cm]{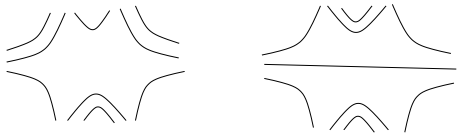}
  \caption{Examples of vertices.}
\label{fig:interactions}
\end{figure}

When evaluating amplitudes of graphs obtained by integrating the last tensor $T$, the tensor network operators act 
as effective vertices (for instance each comes equipped with its own coupling constant). One can 
represent a Feynman graph either as graphs with $D+1$ colors (with the subgraphs with colors $1,\dots,D$, representing the 
effective vertices) or directly as a stranded graph for the tensor $T$.
The latter is built of stranded lines, and vertices representing the connectivity of the indices of the tensors $T$ 
in an effective vertex. For example, for tensors with three indices, both 
\bea
&& T_{a^0 p^1p^2} \bar T_{b^0 p^1 p^2} T_{b^0 q^1 q^2 } \bar T_{c^0 q^1q^2 } T_{c^0 r^1r^2} \bar T_{a^0 r^1r^2}   \; , \crcr
&& T_{ a^0 x^1 b^2 } \bar T_{ p^0 p^1 b^2 }  T_{p^0p^1 c^2  } \bar T_{ d^0 x^1 c^2 } T_{ d^0 q^1q^2} \bar T_{ a^0 q^1q^2 } \; , 
\eea 
are allowed (leading order $\omega(\cB)=0$) vertices. In the representation as stranded graphs 
one represent the index contractions in the vertex 
by strands. Furthermore, listing the indices of the tensors $T$ (respectively $\bar T$) turning clockwise (respectively anticlockwise) 
around a vertex, the two vertices above are represented in figure \ref{fig:interactions}.

The full model of eq. \eqref{eq:genmod} proved for now too complex to be solved analytically. A first step in this direction 
consists in studying simplified models taking into account only subclasses of tensor network operators.
A first subclass is readily identified: one can restrict to vertices such that 
the tensors share alternatively one and $D-1$ indices (the vertex on the left in figure 
 \ref{fig:interactions}). Introducing the shorthand notation $T_{n^1 \vec n}$, where $\vec n = (n^2,\dots n^D)$, they write
\bea
  T_{n^0 \vec n} \bar T_{p^0 \vec n} T_{p^0 \vec n'} \dots  = \Tr{ (TT^{\dagger})^p }
\eea 
where $T$ denoted the $N \times N^{D-1}$ matrix with entries $T_{n^0 \vec n}$. One can check that all such vertices have degree 
$\omega(\cB)=0$. The restricted colored tensor model is then a model of a random $N \times N^{D-1} $ rectangular matrix.
Note that the passage from a stranded graph representation of a Feynman graph to the initial colored representation is
trivial: one needs only to collapse all the stranded lines into colored lines of color $0$, without collapsing the strands in the
vertex. In particular, if a stranded graph is planar, the associated colored graph is also a planar graph. In particular
in $D=3$ a planar colored graph always represents a three sphere\footnote{There are several ways to see this: as it is planar
the graph has trivial homotopy hence it is a sphere, as it is planar it admits a planar jacket
\cite{pcont4,Gur3, GurRiv,GurRiv} and using
\cite{Ryan:2011qm,coloredreview} it is a sphere, etc.}. This is expected to generalize in arbitrary dimensions
\cite{Ryanrest}.

\section{A Matrix model with skewed scalings.} \label{sec:mmodel}

The most general matrix model with skewed scalings for a $N\times N^{D-1}$ matrix $T$ is defined by the partition function 
\bea\label{eq:defmod}
 Z=e^{-N^D F} = \int dT d\bar T e^{-N^{D-1}\Tr V(TT^{\dagger}) } \; ,\qquad V(z) = \sum_{p=1} t_p z^p \; .
\eea

Rectangular matrix models have already been considered in the literature. However, to the best of our knowledge,
this is the first time that a model in which the two indices $n^1$ and $\vec n$ have different scaling in $N$ is considered.
This different scalings have profound consequences on the power counting of graphs (the reader can already infer this by comparing
the $D$ dependent scalings in eq. \eqref{eq:defmod} with those of usual matrix models) 
The rest of this paper is dedicated to solving this model. 

The Feynman graphs generated by \eqref{eq:defmod} are ribbon graphs with two kinds of faces: the ``heavy'' faces carrying an index 
$\vec n$ and the ``light'' faces, carrying an index $n^1$. They can therefore be interpreted as random surfaces decorated by 
patches (corresponding to the light faces). As we will see in the sequel, these light patches completely change the continuum limit of 
the matrix model. Note that scalings in $N$ in eq. \eqref{eq:defmod} are the unique scalings which lead
to a well defined continuum limit.  Most of the analysis we perform below relies on classical matrix models techniques. However, at 
almost every step of the way, the results we obtain are very different from their classical counterparts.
 
The coupling constants $t_p$ can be represented as integrals in the complex plane
\bea \label{eq:couplings}
t_n  = \frac{1}{2\pi \imath} \int_{C} du  \; \frac{1}{u^{n+1}} V(u) \; ,\qquad nt_n = \frac{1}{2\pi \imath} \int_{C} du \;  \frac{1}{u^n} V'(u) \; ,
\eea 
where the contour $C$ is the circle at infinity.

\medskip

{\it Observables} A set of observables is provided by the loop observables, traces of powers of the $N\times N$ matrix $TT^{\dagger}$.
They are computed by deriving w.r.t to the couplings,
\bea
 \frac{1}{N} \frac{ \big \langle \Tr[ (TT^{\dagger})^q ]\big \rangle }{\big\langle 1 \big \rangle}= - N^{-D} \frac{1}{Z}\frac{\partial}{\partial t_q} Z = 
   \frac{\partial}{\partial t_q} F_N \; .
\eea 
We will denote below the generating function of these observables, also known as the resolvent by $W(z)$ and its large $N$ limit by $W_0(z)$
\bea
 W(z) = \frac{1}{z} + \sum_{q=1}^{\infty} \frac{1}{z^{q+1}}
\frac{1}{N}\frac{ \big \langle \Tr[ (TT^{\dagger})^q ]\big \rangle }{\big\langle 1 \big \rangle}
   = \frac{1}{N} \frac{ \Big \langle \Tr \frac{1}{z- TT^{\dagger}} \Big \rangle }{ \big\langle 1 \big \rangle } \; ,
 \qquad W_0(z) = \lim_{N\to \infty} W(z) \; .
\eea 
Remark that the resolvent behaves for $|z| \to \infty $ as $\frac{1}{z}$ for any value of $N$.
The loop observables are computed starting from the resolvent via integrations in the complex plane
\bea
 \frac{1}{N}\frac{ \big \langle \Tr[ (TT^{\dagger})^q ]\big \rangle }{\big\langle 1 \big \rangle} = \frac{1}{2\pi\imath }\int_C du \;  u^q W(u) \; .
\eea 

\medskip

{\it The free energy} $F_N$ is the partition function for connected surfaces. The area of a surface is
proportional with the number of vertices, that is
\bea
 A \sim \frac{1}{F} \sum_{i=2} t_i \partial_{t_i} F \; .
\eea 
This area can be computed as a single derivative.  Renaming the couplings $t_1 = \frac{1}{g}$, $t_i = \frac{\alpha_i}{g}$,  
the potential can be written as $V(z) = \frac{1}{g} \Bigl( z + \sum_{i=2} \alpha_i z^i \Bigr)$
and the derivative of $F_N$ w.r.t. $g$ computes
\bea \label{eq:derivfreerez}
&& g\partial_g F = - N^{-1} \sum_{i=1} t_i \frac{ \Big \langle \Tr[ (MM^{\dagger})^i ]\Big \rangle}{  \Big \langle 1 \Big \rangle  }
= - \frac{1}{2\pi\imath} \int_C du \; V(u)   W(u) \; ,
\eea 
where we used that fact that $V(u)$ is an entire function. It follows that the area of the connected surface is
$A \sim g\partial g \ln F$ (in fact one needs to subtract a finite non universal piece from the free energy prior to taking the logarithm).
Note that for any analytic function $f$
\bea
 \big\langle \Tr f(TT^{\dagger}) \big\rangle = \frac{1}{2\pi\imath}  \int du \;f(u) W(u) \; ,
\eea 
hence solving the model consists in determining  the resolvent $W(u)$.

\subsection{Eigenvalue treatment}

For random matrices one can change variable in the functional integral and pass to $N$ eigenvalue integrals.
This is well known for square matrices and can be done (see \cite{nonhermitian} and references therein) 
also for rectangular ones. In this case one reduces the integral over $T$ and $ T^{\dagger}$ to the 
eigenvalues $\lambda$ of the square $N\times N$ hermitian matrices $TT^{\dagger}$. Using \cite{nonhermitian} we can 
then rewrite the partition function as 
\bea\label{eq:eigenvalues}
 e^{-N^D F_N} = Z_N && =  \int_{0}^{\infty} (\prod_{i=1}^Nd\lambda_i  )
   \prod_{1\le i<j\le N} (\lambda_j -\lambda_i )^2 \; 
  \prod_{i=1}^N  \Bigl( \lambda_i ^{N^{D-1}-N} 
   e^{- N^{D-1} V(\lambda_i)}  \Bigr) \crcr
  &&= \int d\lambda_i e^{-N^{D-1} G (\lambda_i)} \; ,
 \eea 
with 
\bea
 G(\lambda_i) = \sum_i V(\lambda_i)  - \sum_i \ln \lambda_i  
    + \frac{N}{N^{D-1}} \Big{(}  \sum_i \ln \lambda_i  - \frac{2}{N}
   \sum_{i<j} \ln|\lambda_j-\lambda_i| \Big{)} \; .
\eea
The integral in \eqref{eq:eigenvalues} is evaluated by a saddle point whose equations write
\bea\label{eq:saddlev}
 \frac{\partial G}{\partial \lambda_i } =V'(\lambda_i)- \frac{1}{\lambda_i} 
  +  \frac{N}{N^{D-1}} \Big{(} \frac{1}{\lambda_i} -  \frac{2}{N } \sum_{j\neq i} \frac{1}{\lambda_j-\lambda_i} \Big{)} \; .
\eea 
In order to solve at leading order \eqref{eq:saddlev}, one
needs only to note that the last term on the right hand side of eq. \eqref{eq:saddlev} is suppressed in the large
$N$ limit for $D \ge 3$. The saddle point equations therefore {\it decouple} at leading order and the large N free energy is  
\bea\label{eq:soleignevalues}
  F_{\infty} = V(z_0)- \ln z_0\; ,  \quad  z_0 \text{ the physical solution of } z_0 V'(z_0) = \sum_{n=1} n t_n z_0^n=1 \; .
\eea 
The loop observables and resolvent at leading order are therefore 
\bea
&&\lim_{N\to \infty} \frac{1}{N} \frac{\big \langle \Tr[ (TT^{\dagger})^q ]\big \rangle}{ \big\langle 1 \big \rangle }
 = \frac{\partial F_{\infty}}{ \partial t_{q} } 
= z_0^{q} + \Bigl( V'(z_0) - \frac{1}{z_0} \Bigr) \frac{\partial z_0}{\partial t_{q}}= z_0^{q} \;, \crcr
&& W_0(z) = \lim_{N\to \infty} \frac{1}{N} \frac{ \Big \langle \Tr \frac{1}{z- TT^{\dagger}} \Big \rangle }{ \big\langle 1 \big \rangle }
 = \frac{1}{z-z_0} \; .
\eea  
In particular $W_0(z)$ has a {\it pole} singularity at $z_0$, not a cut singularity. This is the first major difference
with the usual matrix models, and a direct consequence of the factorization of the saddle point equations at leading order.

Note that $z_0$ has a straightforward interpretation as the leading order connected two point function of the model, and
the loop observables factor at leading order into two point functions
\bea
 \lim_{N\to \infty} \frac{1}{N} \frac{ \big \langle \Tr[ (TT^{\dagger})^q ]\big \rangle }{\big\langle 1 \big \rangle}
= \Big(   \lim_{N\to \infty} \frac{1}{N} \frac{ \big \langle \Tr[ TT^{\dagger} ]\big \rangle }{\big\langle 1 \big \rangle} \Big)^q
      \; .
\eea 

The large $N$ factorization of the loop observables can readily be understood in terms of graphs.
The graphs contributing to the connected correlation $ \frac{\big \langle \Tr[ (TT^{\dagger})^q ]\big \rangle}{ \big\langle 1 \big \rangle } $
are connected vacuum graph with a marked vertex (corresponding to the insertion $ \Tr[ (TT^{\dagger})^q ] $).
Chose a tree in any graph contributing to this correlation and set its root to be the marked vertex.
Going around the tree one finds that there exists exactly one contraction which closes a maximal number of
faces with ``heavy'' index $\vec n$. A close inspection reveals that this contraction necessarily 
connects any two consecutive $TT^{\dagger}$
on the marked vertex via some connected two point graph, and the factorization follows.

Moreover, as $z_0$ is the two point function, the eq. $z_0 V'(z_0) = 1$ is in fact the Schwinger Dyson Equation (SDE)
\bea
 \frac{1}{Z} \int dT d\bar T \frac{\delta}{ \delta T_{n_1 \vec n} } \Big(  T_{n_1 \vec n} \; e^{-N^{D-1} \Tr V(T\bar T)} \Big) =0 \; ,
\eea 
combined with the factorization property of the loop observables. Also, $z_0 V'(z_0) =1$ is a generating equation for trees
\bea
 z_0 = \frac{1}{t_1} - \sum_{p=2} p \frac{t_p}{t_1} z_0^p = g - \sum_{p=2} p \alpha_p z_0^p \; ,
\eea 
and arbitrary weighted trees exhibit multi critical behaviors. The derivative of free energy $F_{\infty}$ at leading order 
can be computed either directly from \eqref{eq:soleignevalues}  or from \eqref{eq:derivfreerez} and
\bea
 g\partial_g F_{\infty} = -\frac{1}{2\pi \imath} \int_C du \; V(u) \frac{1}{u-z_0} = -V(z_0) \; .
\eea 

\subsection{Multi Critical Points}

Inspired by \cite{Kazakovmulticrit} we choose for potential a polynomial of degree $m$. The saddle point equations 
\bea\label{eq:t1}
 z_0V'(z_0) =1 \Rightarrow g = z_0 + \sum_{q=2}^m q \alpha_q z_0^q \; ,
\eea 
prove that the coupling $g$ is a polynomial in $z_0$. The tensor model achieves its continuum limit when 
\bea
 \frac{\partial g}{\partial z_0} = 0 \; ,
\eea 
and a $m$'th multi critical point when  
\bea
 \frac{\partial g}{\partial z_0} = 0 \dots \frac{\partial^{m-1} g}{(\partial z_0)^{m-1}}  = 0  \; , \qquad 
  \frac{\partial^{m} g}{(\partial z_0)^{m}}  \neq 0 \; .
\eea 
A minimal realization for a $m$'th multi critical model is then obtained if the saddle point equation is
\bea
  g= g_c - (z_c-z_0 )^m \; , \qquad g_c = z_c^m \; ,
\eea 
corresponding to a choice of potential
\bea
 V(z) = \frac{1}{g} z_c^{m} \sum_{q=0}^{m-1} \frac{1}{q+1} + \frac{1}{g} \sum_{q=0}^{m-1} \Bigl( -\frac{z_c^{m-q-1}}{q+1} \Bigr) (z_c-z)^{q+1} \; ,
\qquad z_c = m^{-\frac{1}{m-1}} \; ,
\eea 
where we recall that $V(0)=0$ and the coefficient of $z$ is $\frac{1}{g}$. The leading order resolvent is then
\bea 
 W_0 = \frac{1}{z-z_0} = \sum_{k=0}   \frac{(z_c-z_0)^k }{ (z-z_c)^{k+1} }  = \sum_{k=0} \frac{(g_c-g)^{\frac{k}{m}}}{(z-z_c)^{k+1} } \; ,
\eea
and the leading non analytic behavior of the derivative of the free energy when $g\to g_c$ computes  
\bea
 g\partial_g F_{\infty} \Big{|}_{\rm {n.a.}} \sim  (g_c-g)^{\frac{1}{m}} \Rightarrow F_{\infty} =  (g_c-g)^{ 1 + \frac{1}{m}}
\eea 
corresponding to a susceptibility exponent  $\gamma_m = 1-1/m$,  identical with the one of the multi critical polymers of 
\cite{Ambjorn:1990wp}. This is not surprising, taking into account the earlier remark 
on trees. The $g \to g_c$ limit is a continuum limit as the average area, $A \sim \partial_g \ln F \sim \frac{1}{(g-g_c)}$,
diverges.

\section{Schwinger Dyson equations at all orders} \label{sec:SDE}

In this section we go beyond the leading order and derive the SDE's at all orders. 
We subsequently present the iterative solution order by order in $N$. In order to fit
the scalings in $N$ of various terms one needs to take specific choices for the dimension $D$.
We will do this in the next section.

The derivation of SDE's follows the classical path of \cite{Ambjorn:1990ji,Fukuma:1990jw,Makeenko:1991ry},
up to the unusual scalings of various terms with $N$. 
We reproduce it below for completeness. For $q\ge 1$ we write a SDE
\bea\label{eq:SDE1}
&& \int dT \frac{\delta}{\delta T_{a \vec\alpha}} \Bigl[ \bigl(T (T^{\dagger}T)^q\bigr)_{a \vec \alpha} 
  e^{- N^{D-1}\sum_{p} t_p \Tr[ (TT^{\dagger})^p ] }\Bigl]
\crcr
&&= N \Big\langle \Tr[ (T^{\dagger} T)^q ] \Big \rangle + \sum_{r=1}^{q-1} 
\Big \langle \Tr [ (TT^{\dagger})^{q-r} ]  \Tr [ (T^{\dagger}T)^{r} ] \Big \rangle
 + N^{D-1} \Big\langle \Tr[ (T T^{\dagger} )^q ] \Big \rangle \crcr
&& - N^{D-1} \sum_{p=1} p t_p \Big \langle \Tr[ ( TT^{\dagger}  )^{q+p} ] \Big \rangle =0 \; ,
\eea 
where the double trace term is absent if $q=1$. The case $q=0$ is special and leads to 
\bea\label{eq:SDE2}
N^D \Big \langle 1 \Big \rangle - N^{D-1} \sum_{p} pt_p \Big \langle \Tr[ ( TT^{\dagger}  )^{p} ] \Big \rangle =0 \; .
\eea 

The SDEs can be written as
\bea
  && L_m Z = 0 \; ,\crcr
 && L_q =  - \bigl( N^{-D+2} + 1  \bigr) \frac{\partial}{\partial t_{q} } + N^{-2D+2}\sum_{r=1}^{q-1}  \frac{\partial}{\partial t_{q-r} \partial t_r }
   + \sum_{p=1} p t_p \frac{\partial}{\partial t_{q+p}} \; ,\crcr
 && L_1 =  - \bigl( N^{-D+2} + 1  \bigr) \frac{\partial}{\partial t_{1} }  
   + \sum_{p=1} p t_p \frac{\partial}{\partial t_{p+1}} \; , \crcr
&& L_0 =  N^D + \sum_{p=1} p t_p \frac{\partial}{\partial t_{q+p}} \; ,
\eea 
and, like for the usual matrix models, the $L_m$'s respect the Virasoro algebra.
Introducing the derivative of the resolvent
\bea
 W(z,z) = -  \sum_{p=1}\frac{1}{z^{p }} \frac{d}{dt_p} W(z) \; ,
\eea
a straightforward computations expresses the double trace observable as  
\bea
&& \sum_{t=2} \frac{1}{ z^{t+1} } \sum_{r=1}^{t-1} 
\frac{ \Big \langle \Tr [ (TT^{\dagger})^{t-r} ]  \Tr [ ( T T^{\dagger} )^{r} ] \Big \rangle}
{ \Big\langle 1 \Big \rangle  } \crcr
&&= N^{-D+2} W(z,z) + N^2 \Bigl( zW(z)^2 - 2W(z) + \frac{1}{z} \Bigr) \; .
\eea
Adding the equations \eqref{eq:SDE1} and \eqref{eq:SDE2} with well chosen coefficients leads to 
\bea
&& \Big{[} \frac{1}{z} - \frac{1}{N}  \sum_{p=1} pt_p 
\frac{1}{z}  \frac{  \Big \langle \Tr[ ( TT^{\dagger}  )^{p} ] \Big \rangle }{\Big\langle 1 \Big \rangle }  \Big{]} \crcr
&&+ \bigl( N^{-1}+N^{-D+1} \bigr) \frac{1}{z^2} \frac{  \Big \langle \Tr[ ( TT^{\dagger}  ) ] \Big \rangle }
{\Big\langle 1 \Big \rangle}  -
   \frac{1}{N} \sum_{p=1} p t_p \frac{1}{z^2}
\frac{ \Big \langle \Tr[ ( TT^{\dagger}  )^{1+p} ]  \Big \rangle }{\Big\langle 1 \Big \rangle} \crcr
&&+ \sum_{q=2} \Big{[}  \bigl( N^{-1}+N^{-D+1} \bigr)  \frac{1}{z^{q+1}  }  
\frac{  \Big \langle \Tr[ ( TT^{\dagger}  )^q ] \Big \rangle }{\Big\langle 1 \Big \rangle }    
   +N^{-D} \sum_{r=1}^{q-1} \frac{1}{z^{q+1}}\frac {\Big \langle \Tr [ (TT^{\dagger})^{q-r} ]  \Tr [ (T^{\dagger}T)^{r} ] \Big \rangle}
{\Big\langle 1 \Big \rangle} \crcr
&& - \frac{1}{N}\sum_{p=1} p t_p  \frac{1}{z^{q+1}}
 \frac{ \Big \langle \Tr[ ( TT^{\dagger}  )^{q+p} ] \Big \rangle }{\Big\langle 1 \Big \rangle} 
\Big{]} =0 \; .
\eea 
Recalling that 
\bea
 pt_p = \frac{1}{2\pi\imath} \int_C du \; \frac{1}{u^p} V'(u) \; ,
\eea 
we obtain
\bea
&& \frac{1}{z} + \frac{1}{N} \sum_{q=1} \frac{1}{z^{q+1} } 
\frac{  \Big \langle \Tr[ ( TT^{\dagger}  )^q ] \Big \rangle }{  \Big\langle 1 \Big \rangle }    + 
   N^{-D+1} \sum_{q=1} \frac{1}{z^{q+1} } 
\frac{  \Big \langle \Tr[ ( TT^{\dagger}  )^q ] \Big \rangle }{\Big\langle 1 \Big \rangle  }    \crcr
&&+    N^{-D} \sum_{q=2} \frac{1}{z^{q+1} } 
\sum_{r=1}^{q-1} \frac {\Big \langle \Tr [ (TT^{\dagger})^{q-r} ]  \Tr [ (T^{\dagger}T)^{r} ] \Big \rangle}
{  \Big\langle 1 \Big \rangle   } \crcr
&&  - \frac{1}{N} \sum_{p=1,q=0} 
 \frac{1}{2\pi\imath} \int_C du \; \frac{1}{u^p} V'(u)  \frac{1}{z^{q+1} } 
 \frac{ \Big \langle \Tr[ ( TT^{\dagger}  )^{q+p} ] \Big \rangle }{ \Big\langle 1 \Big \rangle   }  =0 \; ,
\eea 
and a straightforward computations finally yields the loop insertion equation
\bea\label{eq:bub}
 && \frac{1}{2\pi\imath} \int_C du \; \frac{ u V'(u) }{z-u} W(u)  \crcr
  &&= \Bigl( 1-zV'(z) \Bigr) W(z) +
   N^{-D+2} \Bigl(  zW(z)^2  - W(z) \Bigr) + N^{-2D+2} W(z,z) \; .
\eea 
Note that for $D=2$ we obtain the usual loop insertion equation for a non hermitian matrix (see for 
instance \cite{Ambjorn:1990ji}). The loop insertion equation simplifies further.
Indeed, at leading order $W_0(z)$ depends on $t_i$ only implicitly trough $z_0$. This generalizes order by order
to $W(z)$. Noting that 
\bea
 z_0V'(z_0) - 1 =0 \Rightarrow \Bigl[ z V'(z) \Bigr]'_{z=z_0} dz_0 + \sum_{n=1}n z_0^n dt_n=0 
 \Rightarrow \frac{\partial z_0}{\partial t_q} = - \frac{ q z_0^{q} }{ \Bigl[ z V'(z) \Bigr]'_{z=z_0} } \; ,
\eea 
where we supposed that $\Bigl[ z V'(z) \Bigr]'_{z=z_0} \neq 0$ (which holds for instance for multi critical models),
we can evaluate   
\bea
 W(z,z) = - \sum_{q=1} \frac{1}{z^{q} } \frac{\partial}{\partial t_q} W(z) = 
     \partial_{z_0} W(z) \sum_{q=1} \frac{1}{z^q} \frac{ z_0 \frac{d}{dz_0} z_0^q }{  \Bigl[ z V'(z) \Bigr] '\Big{|}_{z=z_0}   } \; ,
\eea 
and for $z>z_0$ we obtain 
\bea
 W(z,z) = \frac{ z z_0 \; \partial_{z_0} W(z)}{  (z-z_0)^2   \Bigl[ z V'(z) \Bigr] '\Big{|}_{z=z_0}      } \; ,
\eea 
and the equation \eqref{eq:bub} becomes  
\bea \label{eq:bubbgenfin}
\frac{1}{2\pi\imath} \int_C du \; \frac{ u V'(u) }{z-u} W(u) &=& \Bigl( 1-zV'(z) \Bigr) W(z) +
   N^{-D+2} \Bigl(  zW(z)^2  - W(z) \Bigr) \crcr
  &&+ N^{-2D+2}  \frac{ z z_0 \; \partial_{z_0} W(z)}{  (z-z_0)^2   \Bigl[ z V'(z) \Bigr] '\Big{|}_{z=z_0}      } \; .
\eea 

\subsection{Iterative Solution}

The loop insertion equation \eqref{eq:bubbgenfin} can be solved order by order in $N$.
We start by expanding the resolvent in powers of $N$,
\bea
  W(z) = W_0 (z) + \sum_{n}N^{-n} W_n(z) \; .
\eea 
From eq. \eqref{eq:bubbgenfin} 
we conclude that the first non trivial correction is $W_{D-2}$, and it respects the equation
\bea
 \frac{1}{2\pi\imath} \int_C du \; \frac{ u V'(u) } {z-u} W_{D-2}(u) - \Bigl( 1-zV'(z) \Bigr) W_{D-2}(z) 
   = z W_0^2 - W_0 = \frac{z_0}{(z-z_0)^2} \; .
\eea 
Note that 
\bea
&& \frac{1}{2\pi\imath} \int_C du \; \frac{ u V'(u) } {z-u}  \frac{ 1 }{(u-z_0)^{n+1}}  - 
\Bigl( 1-zV'(z) \Bigr)  \frac{ 1 }{(z-z_0)^{n+1}} \crcr
&&= \sum_{p=0}^n \frac{1}{ (z-z_0)^{p+1}  } \frac{1}{(n-p)!} \Bigl[ uV'(u) \Bigr]_{z_0}^{(n-p)} - 
 \frac{1}{(z-z_0)^{n+1}} \crcr
&&= \sum_{p=0}^{n-1} \frac{1}{ (z-z_0)^{p+1}  } \frac{1}{(n-p)!} \Bigl[ uV'(u) \Bigr]_{z_0}^{(n-p)} \; ,
\eea 
where we took into account $z_0V'(z_0)=1$. Proposing the ansatz
\bea
 W_{D-2} =  \frac{a}{(z-z_0)^3} + \frac{b}{(z-z_0)^2} \; ,
\eea 
The equation writes
\bea
 a \Bigl(\frac{1}{(z-z_0)^2}\Bigl[ uV'(u) \Bigr]'_{z_0} + \frac{1}{z-z_0}  \frac{1}{2}\Bigl[ uV'(u) \Bigr]''_{z_0} \Bigr)
 + b\Bigl(\frac{1}{z-z_0}  \Bigl[ uV'(u) \Bigr]'_{z_0}  \Bigr) = \frac{z_0}{(z-z_0)^2} \;,
\eea 
with solution
\bea
 a = \frac{z_0}{  \Bigl[ uV'(u) \Bigr]'_{z_0}  } \qquad b= - \frac{ z_0 \Bigl[ uV'(u) \Bigr]''_{z_0}  }{2 
\Big{\{}\Bigl[ uV'(u) \Bigr]'_{z_0}  \Big{\}}^2 } \; .
\eea 

The next non trivial term is $W_{2D-4}$. Subsequent terms are $W_{3D-6}$ and $W_{2D-2}$ (generated by the last line in 
eq. \eqref{eq:bubbgenfin}). We obtain non trivial corrections $W_n$ for $n = p(D-2) +q(2D-2) $. In order to
derive the full solution we will in the sequel chose $D=3$.

\section{The double scaling limit in $D=3$}\label{sec:multicrit}

In $D=3$ we obtain a nonzero correction $W_n$ for all $n$ 
\bea
 W(z) = W_0(z)+\sum_{i=1}N^{-n}W_n(z) \; .
\eea 
Note that for any even $D$ (both $D=4$ and in $D=2$ for instance) nontrivial corrections $W_n$ arise only for even $n$.  
The equation \eqref{eq:bubbgenfin} translates for $W_n$ into
\bea 
&&\frac{1}{2\pi\imath} \int_C du \; \frac{ u V'(u) }{z-u} W_n(u)  -  \Bigl( 1-zV'(z) \Bigr) W_n(z) \crcr
&&= z  \sum_{p=0}^{n-1} W_{p}(z) W_{n-1-p}(z) - W_{n-1}(z)
+ \frac{ z z_0 \; }{  (z-z_0)^2   \Bigl[ uV'(u) \Bigr]'_{z_0}     } \partial_{z_0} W_{n-4}(z) \; ,
\eea 
where by convention $W_q = 0$ for $q<0$. To solve the equation for $W_n$, we propose the ansatz
\bea
  W_n(z) = \sum_{k=1}^{2n} \frac{f_n^k(z_0)}{(z-z_0)^{k+1}}  \quad n\ge 1 \; , \qquad W_0  = \frac{1}{z-z_0} \; ,
\eea 
and substituting we obtain for $n\ge 1$
\bea
&& \sum_{p=1}^{2n} f_n^p(z_0)  \sum_{k=0}^{p-1} \frac{1}{ (z-z_0)^{k+1}  } \frac{  1}{(p-k)!} \Bigl[ uV'(u) \Bigr]_{z_0}^{(p-k)}  \crcr
&&= (z-z_0)\sum_{p=0}^{n-1} W_{p}(z) W_{n-1-p}(z) + z_0 \sum_{p=0}^{n-1} W_{p}(z) W_{n-1-p}(z)
   - W_{n-1}(z) \crcr
&& + \frac{ z_0 \; }{  (z-z_0) \Bigl[ uV'(u) \Bigr]'_{z_0}       } \partial_{z_0} W_{n-4}(z)
  +\frac{  z_0^2 \; }{  (z-z_0)^2  \Bigl[ uV'(u) \Bigr]'_{z_0}     } \partial_{z_0} W_{n-4}(z) \; .
\eea 
Exchanging the sums on the left hand side we obtain
\bea \label{eq:finfin}
&& \sum_{k=0}^{2n-1} \frac{1}{ (z-z_0)^{k+1}  } \sum_{p=k+1}^{2n} f_n^p(z_0)   \frac{  1}{(p-k)!} \Bigl[ uV'(u) \Bigr]_{z_0}^{(p-k)}  \crcr
&&=  z_0 \sum_{p=0}^{n-1} W_{p}(z) W_{n-1-p}(z) + (z-z_0)\sum_{p=0}^{n-1} W_{p}(z) W_{n-1-p}(z) 
   - W_{n-1}(z) \crcr
&& + \frac{ z_0 \; }{  (z-z_0) \Bigl[ uV'(u) \Bigr]'_{z_0}       } \partial_{z_0} W_{n-4}(z)
  +\frac{  z_0^2 \; }{  (z-z_0)^2  \Bigl[ uV'(u) \Bigr]'_{z_0}     } \partial_{z_0} W_{n-4}(z)  \; .
\eea 

The equation \eqref{eq:finfin} allows to determine iteratively the coefficients $ f_n^p(z_0)  $. 
In order to prove that a multi critical model admits a double scaling limit we are interested only in the
behavior of these coefficients when $z_0$ approaches $z_c$. Using the solution at first order we see
that at leading order in $z_c-z_0$,
\bea
 f^2_1 (z_0) = \frac{ d^2_1 }{ (z_c-z_0)^{m-1} } \; , \qquad  f^1_1(z_0) = \frac{d^1_1}{ (z_c-z_0)^m } \; ,
\eea 
with $ d^2_1 $ and $ d^1_1 $ some numerical coefficients. We will show by induction that the generic scaling 
formula
\bea \label{eq:scalscal}
  f_n^k(z_0)= \frac{ d_n^k}{ (z_c-z_0)^{ n (m +1) - k     }   } \; ,
\eea 
holds. The scaling at order $n$ is fixed from \eqref{eq:finfin}. We analyze one by one the scalings of the various terms 
on the right hand side for $n\ge 2$

The first term we analyze is
\bea
   \sum_{p=0}^{n-1} W_{p}(z) W_{n-1-p}(z) = 2W_0W_{n-1} + \sum_{p=0}^{n-2} W_{p}(z) W_{n-1-p}(z) \; ,
\eea 
where the second sum is absent when $n=2$. It writes
\bea
&&  2 W_0 W_{n-1}  + \sum_{p=1}^{n-2} \sum_{k_1=1}^{2p} \sum_{k_2=1}^{2n-2-2p} \frac{f_{p}^{k_1} (z_0) }{ (z-z_0)^{k_1+1} } 
\frac{ f_{n-1-p}^{k_2}(z_0) }{ (z-z_0)^{k_2+1} } \crcr
&& = \sum_{t=1}^{2n-2} \frac{2 f_{n-1}^t(z_0)}{(z-z_0)^{t+2}} +
 \sum_{p=0}^{n-2} \sum_{k_1=1}^{2p} \sum_{t=2 }^{2n-2} \frac{1}{(z-z_0)^{ t + 2 }} f_{p}^{k_1} (z_0) f_{n-1-p}^{t-k_1}(z_0) \crcr
&&\sim \sum \frac{1}{(z-z_0)^{k+1}}  \frac{1}{(z_c-z_0)^{(n-1)(m+1) -k+1 }} \; ,
\eea 
hence the first line of the right hand side of \eqref{eq:finfin} is dominated when $z_0\to z_c$ by the first term.
The derivatives of $W_{n-4}$ behave for $z_0\to z_c$ like
\bea
 \partial_{z_0}W_{n-4} \sim   \sum_{k} \frac{1}{(z-z_0)^{k+1}} \frac{1}{ (z_c-z_0)^{(n-4)(m+1) - k +1 } } \; ,
\eea
hence the second line on the right hand side of \eqref{eq:finfin} scales at most like
\bea
  \sum_{k} \frac{1}{(z-z_0)^{k+1}} \frac{1}{ (z_c-z_0)^{(n-3)(m+1) - k +3 } } \; ,
\eea 
and when $z_0\to z_c$ is dominated by the terms in the first line. Identifying the coefficients of $ \frac{1}{(z-z_0)^{k+1} } $
in eq. \eqref{eq:finfin} we conclude that 
\bea
 \sum_{p=k+1}^{2n} f_{n}^p (z_0) [uV']^{(p-k)} \sim  \frac{1}{ (z_c-z_0)^{ (n-1) (m+1) -  k  +1 }   } \; ,
\eea
and taking into account that $[uV']^{(p-k)}\sim (z_c-z_0)^{m-p+k}  $, we get 
\bea
 f_{n}^p (z_0) \sim \frac{1}{ (z_c-z_0)^{ n m + n - p     }   } \; ,
\eea 
reproducing \eqref{eq:scalscal}. It follows that the leading contributions to the resolvent write 
\bea
 W(z)&& = \frac{1}{z-z_0} + \sum_{n=1}N^{-n} \sum_{k=1}  \frac{ (z_c-z_0)^k }{ (z-z_0)^{k+1}  }  \frac{d^k_n}{ (z_c-z_0)^{ n (m + 1) }   } \crcr
&& =  \frac{1}{z-z_0} + \sum_{k=1}   \frac{ (z_c-z_0)^k }{ (z-z_0)^{k+1}  }  \sum_{n}
   d^k_n \frac{1}{ [N(z_c-z_0)^{m+1}]^n } \; .
\eea 
Taking into account that $z_c-z_0 = (g_c -g)^{\frac{1}{m}}$, the resolvent writes when $g\to g_c$ as
\bea
W(z) =  \frac{1}{z-z_c} + \sum_k \frac{1}{ (z-z_c)^{k+1} }  (g_c-g)^\frac{k}{m} F_{k}(\kappa)
\eea 
with $\kappa^{-1} = N(g_c-g_0)^{1+\frac{1}{m}}$. 
The graphs contributing to the double scaling limit can be easily understood in terms of stranded graphs 
starting from the SDE in equation 
\eqref{eq:bub}. The term on the left hand side represents the connection of two effective vertices by a tree line. 
The terms on the left hand side represent a marked vertex decorated by a loop line. The loop line divides the
marked vertex into two smaller effective vertices, corresponding to 
$ \frac{1}{ \big \langle 1  \big \rangle } \big \langle \Tr [ (TT^{\dagger})^{q-r} ]  \Tr [ (T^{\dagger}T)^{r} ] \big \rangle $.
Either these two smaller vertices
do not reconnect again (corresponding to the terms involving $W(z)$ and $W(z)^2$), i.e. the 
correlation splits into two connected correlations
$\big \langle \Tr [ (TT^{\dagger})^{q-r} ] \big\rangle_c \big \langle \Tr [ (T^{\dagger}T)^{r} ] \big \rangle_c $, 
in which case the total graph is planar. Or the two smaller vertices reconnect with at least a line (the terms generated by $W(z,z)$), 
in which case the correlation does not split $ \big \langle \Tr [ (TT^{\dagger})^{q-r} ]  \Tr [ (T^{\dagger}T)^{r} ] \big \rangle_c $,
and the initial loop line and the new line cross yielding a non planar graph.
All the contributions of $W(z,z)$ are suppressed in the double scaling limit with respect to those of $W^2(z)$.
It ensues that only planar graphs corresponding to spheres in $D=3$ dimension contribute in the double scaling
limit. 

\section{Conclusion} \label{sec:conc}

We have shown that a $N\times N^{D-1}$ matrix model admits multi critical points and a 
double scaling limit. The model constitutes a toy model for higher dimensional tensor models. In $D=3$ 
the double scaling limit only graphs having the topology of a sphere contribute.
This is expected to hold for all $D$. Indeed, form eq. \eqref{eq:bub} we see that the term $W(z,z)$ is increasingly suppressed
with respect to $W^2(z)$, hence only planar graphs should dominate for all $D$. Such graphs are homotopically 
trivial in all dimensions, and are expected to represent spheres.

Although the model we analyze is very simple, and the (prospective) double scaling limit of colored tensor models is expected to 
be much more involved one can draw several conclusions about the major features of such an expansion. First, one should not 
expect all terms in the $1/N$ expansion to contribute: unlike in usual matrix models, even in the double scaling many terms 
are suppressed. In fact, somewhat surprisingly, one should not expect the double scaling limit to be a summation over topologies:
it seems possible that only spherical graphs contribute. Third, the exact interpretation of this double scaling limit is yet unclear.
In two dimensions one notices that the free energy in the double scaling limit is a series in $\kappa^{\chi}$, where
$\chi$ is the Euler character of a graph. As the latter is just the evaluation of the Einstein Hilbert action, one 
concludes that the double scaling limit corresponds to a continuum limit with a finite (renormalized) Newton's constant.
In higher dimension the precise interpretation of the parameter $\kappa$ is far from clear.

\section*{Acknowledgements}

Research at Perimeter Institute is supported by the Government of Canada through Industry
Canada and by the Province of Ontario through the Ministry of Research and Innovation.

\end{document}